# An Approximation Algorithm for the Link Building Problem[*]

Martin Olsen[1], Anastasios Viglas[2], and Ilia Zvedeniouk[2]

[1] AU Herning, Aarhus University,
Birk Centerpark 15, DK-7400 Herning, Denmark
`martino@hih.au.dk`
[2] University of Sydney, 1 Cleveland St, NSW 2006, Australia
`taso.viglas@sydney.edu.au, izve6419@uni.sydney.edu.au`

**Abstract.** In this work we consider the problem of maximizing the PageRank of a given target node in a graph by adding $k$ new links. We consider the case that the new links must point to the given target node (backlinks). Previous work shows that this problem has no fully polynomial time approximation schemes unless $P = NP$. We present a polynomial time algorithm yielding a PageRank value within a constant factor from the optimal. We also consider the naive algorithm where we choose backlinks from nodes with high PageRank values compared to the outdegree and show that the naive algorithm performs much worse on certain graphs compared to the constant factor approximation scheme.

## 1 Introduction

Search engine optimization (SEO) is a fast growing industry that deals with optimizing the ranking of web pages in search engine results. SEO is a complex task, especially since the specific details of search and ranking algorithms are often not publicly released, and also can change frequently. One of the key elements of optimizing for search engine visibility is the "external link popularity", which is based on the structure of the web graph. The problem of obtaining optimal new backlinks in order to achieve good search engine rankings is known as Link Building and leading experts from the SEO industry consider Link Building to be an important aspect of SEO [5].

The PageRank algorithm is one of the more popular methods of defining a ranking among nodes according to the link structure of the graph. The definition of PageRank [3] uses random walks based on the random surfer model. The random surfer walk is defined as follows: the walk can start from any node in the graph and at each step the surfer chooses a new node to visit. The surfer "usually" chooses (uniformly at random) an outgoing link from the current node, and follows it. But with a small probability at each step the surfer may choose to ignore the current node's outgoing links, and just zap to any node in the graph

---

(chosen uniformly at random). The purpose of zapping is to make all nodes in the graph reachable, and to make the start of the walk irrelevant (for the purposes of PageRank). The random surfer walk is a random walk on the graph with random restarts every few steps. This random walk has a unique stationary probability distribution that assigns the probability value $\pi_i$ to node $i$. This value is the PageRank of node $i$, and can be interpreted as the probability for the random surfer of being at node $i$ at any given point during the walk. We refer to the random restart as zapping. The parameter that controls the zapping frequency is the probability of continuing the random walk at each step, $\alpha > 0$. The high level idea is that the PageRank algorithm will assign high PageRank values to nodes that would appear more often in a random surfer type of walk. In other words, the nodes with high PageRank are hot-spots that will see more random surfer traffic, resulting directly from the link structure of the graph. If we add a small number of new links to the graph, the PageRank values of certain nodes can be affected very significantly. The Link Building problem arises as a natural question: given a specific target node in the graph, what is the best set of $k$ links that will achieve the maximum increase for the PageRank of the target node?

We consider the problem of choosing the optimal set of *backlinks* for maximizing $\pi_x$, the PageRank value of some target node $x$. A backlink (with respect to a target node $x$) is a link from any node towards $x$. Given a graph $G(V, E)$ and an integer $k$, we want to identify the $k \geq 1$ links to add to node $x$ in $G$ in order to maximize the resulting PageRank of $x$, $\pi_x$. Intuitively, the new links added should redirect the random surfer walk towards the target node as often as possible. For example, adding a new link from a node of very high PageRank would usually be a good choice.

### 1.1 Related Work and Contribution

The PageRank algorithm [3] is based on properties of Markov chains. There are many results related to the computation of PageRank values [6,2] and re-calculating PageRank values after adding a set of new links in a graph [1].

The Link Building problem that we consider in this work is known to be NP-hard [9] where it is even showed that there is no fully polynomial time approximation scheme (FPTAS) for Link Building unless NP = P and the problem is also shown to be W[1]-hard with parameter k. A related problem considers the case where a target node aims at maximizing its PageRank by adding new outlinks. Note that in this case, new outlinks can actually decrease the PageRank of the target node. This is different to the case of the Link Building problem with backlinks where the PageRank of the target can only increase [1]. For the problem of maximizing PageRank with outlinks we refer to [1,4] containing, among other things, guidelines for optimal linking structure.

In Sect. 2 we give background to the PageRank algorithm. In Sect. 3 we formally introduce the Link Building problem. In Sect. 3.1 we consider the naive and intuitively clear algorithm for Link Building where we choose backlinks from the nodes with the highest PageRank values compared to their outdegree (plus one). We show how to construct graphs where we obtain a surprisingly high approximation ratio. The approximation ratio is the value of the optimal solution divided by the value of the solution obtained by the algorithm. In Sect. 3.2, we present a polynomial time algorithm yielding a PageRank value within a constant factor from the optimal and therefore show that the Link Building problem is in the class APX.

## 2  Background: The PageRank Algorithm

The PageRank algorithm was proposed by Brin, Page and Brin and Page [3] as a webpage ranking method that captures the importance of webpages. Loosely speaking, a link pointing to a webpage is considered a vote of trust for that webpage. A link from an important webpage is better for the receiver than a link from an unimportant webpage.

We consider directed graphs $G = (V, E)$ that are unweighted and therefore we count multiple links from a node $u$ to a node $v$ as a single link. The graph may represent a set of webpages $V$ with hyperlinks between them, $E$, or any other linked structure.

We define the following *random surfer* walk on $G$: at every step the random surfer will choose a new node to visit. If the random surfer is currently visiting node $u$ then the next node is chosen as follows: (1) with probability $\alpha$ the surfer chooses an outlink from $u$, $(u, v)$, uniformly at random and visits $v$. If the current node $u$ happens to be a sink (and therefore has no outlinks) then the surfer picks any node $v \in V$ uniformly at random, (2) with probability $1 - \alpha$ the surfer visits any node $v \in V$ chosen uniformly at random– this is referred to as *zapping*. A typical value for the probability $\alpha$ is 0.85. The random surfer walk is therefore a random walk that usually follows a random outlink, but every few steps it essentially restarts the random walk from a random node in the graph.

Since the new node depends only on the current position in the graph, the sequence of visited pages is a Markov chain with state space $V$ and transition probabilities that can be defined as follows. Let $P = \{p_{ij}\}$ denote a matrix derived from the adjacency matrix of the graph $G$, such that $p_{ij} = \frac{1}{\text{outdeg}(i)}$ if $(i, j) \in E$ and 0 otherwise (outdeg$(i)$ denotes the outdegree of $i$, the number of out-going edges from node $i \in V$). If outdeg$(i) = 0$ then $p_{ij} = \frac{1}{n}$. The transition probability matrix of the Markov chain that describes the random surfer walk can therefore be written as $Q = \frac{1-\alpha}{n} \mathbb{1}_{n,n} + \alpha P$, where $\mathbb{1}_{n,n}$ is an $n \times n$ matrix with every entry equal to 1.

This Markov chain is aperiodic and irreducible and therefore has a unique stationary probability distribution $\pi$ - the eigenvector associated with the dominant eigenvalue of $Q$. For any positive initial probability distribution $x_0$ over $V$, the iteration $x_0^T Q^l$ will converge to the stationary probability distribution $\pi^T$ for large enough $l$. This is referred to as the *power method* [6].

The distribution $\pi = (\pi_1, \ldots, \pi_n)^T$ is defined as the PageRank vector of $G$. The PageRank value of a node $u \in V$ is the *expected* fraction of visits to $u$ after $i$ steps for large $i$ regardless of the starting point. A node that is reachable from many other nodes in the graph via short directed paths will have a larger PageRank, for example.

## 3 The Link Building Problem

The $k$ backlink (or Link Building) problem is defined as follows:

**Definition 1.** *The* LINK BUILDING *problem:*

- *Instance: A triple $(G, x, k)$ where $G(V, E)$ is a directed graph, $x \in V$ and $k \in \mathbb{Z}^+$.*
- *Solution: A set $S \subseteq V \setminus \{x\}$ with $|S| = k$ maximizing $\pi_x$ in $G(V, E \cup (S \times \{x\}))$.*

For fixed $k = 1$ this problem can be solved in polynomial time by simply calculating the new potential PageRanks of the target node after adding a link from each node. This requires $O(n)$ PageRank calculations. The argument is similar for any fixed $k$. As mentioned in Sect. 1.1, if $k$ is part of the input then the problem becomes NP-hard.

### 3.1 Naive Selection of Backlinks

When choosing new incoming links in a graph, based on the definition of the PageRank algorithm, higher PageRank nodes appear to be more desirable. If we naively assume that the PageRank values will not change after inserting new links to the target node then the optimal new sources for links to the target would be the nodes with the highest PageRank values compared to outdegree plus one. This leads us to the following naive but intuitively clear algorithm:

**Naive**$(G, x, k)$
 Compute all PageRanks $\pi_i$, for all $(i \in V : (i, x) \notin E)$
 Return the $k$ webpages with highest values of $\frac{\pi_i}{d_i + 1}$, where $d_i$ is the outdegree of page $i$

**Fig. 1.** The naive algorithm

The algorithm simply computes all initial PageRanks and chooses the $k$ nodes with the highest value of $\frac{\pi_i}{d_i + 1}$. It is well understood [8] that the naive algorithm is not always optimal. We will now show how to construct graphs with a surprisingly high approximation ratio – roughly 13.8 for $\alpha = 0.85$ – for the naive algorithm.

**Lower Bound for the Approximation Ratio of the Naive Algorithm**
We define a family of input graphs ("cycle versus sink" graphs) that have the following structure: There is a cycle with $k$ nodes, where each node has a number of incoming links from $t_c$ other nodes (referred to as *tail nodes*). Tail nodes are used to boost the PageRanks of certain pages in the input graph and have an outdegree of 1. There are also $k$ sink nodes (no outlinks) each one with a tail of $t_s$ nodes pointing to them. The target node is $x$ and it has outlinks towards all of the sinks. Figure 2 illustrates this family of graphs. Assume also that there is an isolated large clique with size $t_i$. The purpose of this clique is essentially to absorb all the zapping traffic. Intuitively, this makes the linking structural elements more important. Later we also give the bound without this clique, and see that it is worse.

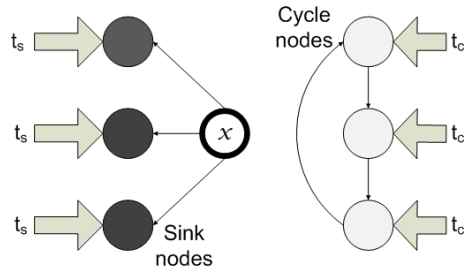

**Fig. 2.** A "cycle versus sink" graph for the naive algorithm.

Due to symmetry all pages in the cycle will have the same PageRank $\pi_c$ and the $k$ sink pages will have the PageRank $\pi_s$. All tail nodes have no incoming links and, due to symmetry, will have the same PageRank denoted by $\pi_t$. The PageRank of the target node is $\pi_x$ and the PageRank of each node in the isolated clique is $\pi_i$.

The initial PageRanks for this kind of symmetric graph can be computed by writing a linear system of equations based on the identity $\pi^T = \pi^T Q$. The total number of nodes is $n = k\,(t_s + t_c + 2) + t_i + 1$.

$$\pi_t = \frac{1-\alpha}{n} + \frac{\alpha\,k\,\pi_s}{n}$$
$$\pi_x = \pi_t = \frac{1-\alpha}{n} + \frac{\alpha\,k\,\pi_s}{n}$$
$$\pi_s = \pi_t + \alpha\left(\frac{\pi_x}{k} + t_s \pi_t\right)$$
$$\pi_c = \pi_t + \alpha\left(\pi_c + t_c \pi_t\right)$$
$$\pi_i = \pi_t + \alpha \pi_i\ .$$

We need to add $k$ new links towards the target node. We will pick the sizes of the tails $t_c$, $t_s$ and therefore the PageRanks in the initial network so that the PageRank (divided by outdegree plus one) of the cycle nodes is slightly higher than the PageRank over degree for the sink nodes. Therefore the naive algorithm 1 will choose to add $k$ links from the $k$ cycle nodes. Once one link has been added, the rest of the cycle nodes are not desirable anymore, a fact that the naive algorithm fails to observe. The optimal solution is to add $k$ links from the sink nodes, as each node in a sense directs *independent* traffic to the target.

In order to make sure cycle nodes are chosen by the naive algorithm, we need to ensure that $\frac{\pi_c}{outdeg(c)+1} > \frac{\pi_s}{outdeg(s)+1} \Leftrightarrow \frac{\pi_c}{2} > \pi_s \Leftrightarrow \pi_c/\pi_s = 2 + \delta$ for some $\delta > 0$. We then parameterize our tails:

$$t_c = u \tag{1}$$
$$t_i = u^2 \tag{2}$$
$$t_s = \frac{u}{2(1 - \lambda\alpha)} \; . \tag{3}$$

where $u$ determines the size of the graph and $\lambda$ is the solution of $\pi_c/\pi_s = 2+\delta$, giving

$$\lambda = \frac{\left((\alpha^2 - \alpha)\delta + 2\alpha^2\right)ku + 2((\alpha-1)\delta + 2\alpha - 1)k + 2(\alpha^2 - \alpha)\delta + 4(\alpha^2 - \alpha)}{2\alpha^2 ku + ((2\alpha^2 - 2\alpha)\delta + 4\alpha^2 - 2\alpha)k + (2\alpha^3 - 2\alpha^2)\delta + 4\alpha^3 - 4\alpha^2}$$

We can solve for $\lambda$ for any desired value of $\delta$. Note also that we choose the tails of the clique nodes to be $u^2$ in order to make them asymptotically dominate all the other tails. The naive algorithm therefore will add $k$ links from the cycle nodes which will result in the following linear system for the PageRanks:

$$\pi_t^g = \frac{1-\alpha}{n} + \frac{\alpha k \pi_s^g}{n}$$
$$\pi_x^g = \pi_t^g + \alpha k \frac{\pi_c^g}{2}$$
$$\pi_s^g = \pi_t^g + \alpha \left(\frac{\pi_x^g}{k} + t_s \pi_t^g\right)$$
$$\pi_c^g = \pi_t^g + \alpha \left(\pi_c^g/2 + t_c \pi_t^g\right)$$
$$\pi_i^g = \pi_t^g + \alpha \pi_i^g \; .$$

The optimal is to choose $k$ links from the sink nodes with a resulting Page-Rank vector described by the following system:

$$\pi_t^o = \frac{1-\alpha}{n}$$
$$\pi_x^o = \pi_t^o + \alpha k \pi_s^o$$
$$\pi_s^o = \pi_t^o + \alpha \left( \frac{\pi_x^o}{k} + t_s \pi_t^o \right)$$
$$\pi_c^o = \pi_t^o + \alpha \left( \pi_c^o + t_c \pi_t^o \right)$$
$$\pi_i^o = \pi_t^o + \alpha \pi_i^o \ .$$

We solve these systems and calculate the approximation ratio of the naive algorithm:

$$\frac{\pi_x^o}{\pi_x^g} = \frac{\left(\alpha^3 - 2\alpha^2\right) k\, t_s + \left(\alpha^2 - 2\alpha\right) k + \alpha - 2}{\left(\alpha^4 - \alpha^2\right) k\, t_c + \left(\alpha^3 - \alpha\right) k - \alpha^3 + 2\alpha^2 + \alpha - 2} \ . \tag{4}$$

We now set our tails as described above in Equations 1-3 and let $u, k \to \infty$. So for large values of the tail sizes we get the following limit:

$$\lim_{u,k \to \infty} \frac{\pi_x^o}{\pi_x^g} = \frac{2 - \alpha}{\left(\alpha^3 - \alpha^2 - \alpha + 1\right) \delta + 2\alpha^3 - 2\alpha^2 - 2\alpha + 2} \ . \tag{5}$$

Now letting $\delta \to 0$ (as any positive value serves our purpose) we get the following theorem.

**Theorem 1.** *Consider the Link Building problem with target node $x$. Let $G = (V, E)$ be some directed graph. Let $\pi_x^o$ denote the highest possible PageRank that the target node can achieve after adding $k$ links, and $\pi_x^g$ denote the PageRank after adding the links returned by the naive algorithm from Fig. 1. Then for any $\epsilon > 0$ there exist infinitely many different graphs $G$ where*

$$\frac{\pi_x^o}{\pi_x^g} > \frac{2 - \alpha}{2(1-\alpha)(1-\alpha^2)} - \epsilon \ . \tag{6}$$

Note that $\epsilon$ can be written as function of $u$, $\delta$, $k$ and $\alpha$. As $u, k \to \infty$, $\epsilon \to 0$ giving the asymptotic lower bound. For $\alpha = 0.85$ the lower bound is about 13.8.

To see that the large isolated clique is necessary, we follow the same procedure as above but setting $t_i = 0$, which gives us the inferior bound

$$-\frac{2\alpha^4 + \alpha^2 + \alpha - 6}{4\alpha^3 - 6\alpha^2 - 4\alpha + 6}$$

which is only about 4.7 for $\alpha = 0.85$.

### 3.2 Link Building is in APX

In this section we present a greedy polynomial time algorithm for Link Building; computing a set of $k$ new backlinks to target node $x$ with a corresponding value of $\pi_x^G$ within a constant factor from the optimal value. In other words we prove that Link Building is a member of the complexity class APX. We also introduce $z_{ij}$ as the expected number of visits of node $j$ starting at node $i$ without zapping within the random surfer model. These values can be computed in polynomial time [1].

**Proof of APX Membership** Now consider the algorithm consisting of $k$ steps where we at each step add a backlink to node $x$ producing the maximum increase in $\frac{\pi_x}{z_{xx}}$ – the pseudo code of the algorithm is shown in Fig. 3. This algorithm runs in polynomial time, producing a solution to the Link Building problem within a constant factor from the optimal value as stated by the following theorem. So, Link Building is a member of the complexity class APX.

>   **r-Greedy**$(G, x, k)$
>     $S := \emptyset$
>     **repeat** $k$ **times**
>       Let $u$ be a node which maximizes the value of $\frac{\pi_x}{z_{xx}}$ in $G(V, E \cup \{(u, x)\})$
>       $S := S \cup \{u\}$
>       $E := E \cup \{(u, x)\}$
>     Report $S$ as the solution

**Fig. 3.** Pseudo code for the greedy approach.

**Theorem 2.** *We let $\pi_x^G$ and $z_{xx}^G$ denote the values obtained by the r-Greedy algorithm in Fig. 3. Denoting the optimal value bye $\pi_x^o$, we have the following*

$$\pi_x^G \geq \pi_x^o \frac{z_{xx}^G}{z_{xx}^o}(1 - \frac{1}{e}) \geq \pi_x^o (1 - \alpha^2)(1 - \frac{1}{e}) \ .$$

*where $e = 2.71828\ldots$ and $z_{xx}^o$ is the value of $z_{xx}$ corresponding to $\pi_x^o$.*

*Proof.* Proposition 2.1 in [1] by Avrachenkov and Litvak states the following

$$\pi_x = \frac{1-\alpha}{n} z_{xx} (1 + \sum_{i \neq x} r_{ix}) \ . \qquad (7)$$

where $r_{ix}$ is the probability that a random surfer starting at $i$ reaches $x$ before zapping. This means that the algorithm in Fig. 3 greedily adds backlinks to $x$ in an attempt to maximize the probability of reaching node $x$ before zapping, for a surfer dropped at a node chosen uniformly at random. We show in Lemma 1 below that $r_{ix}$ in the graph obtained by adding links from $X \subseteq V$ to $x$ is a *submodular* function of $X$ – informally this means that adding the link $(u, x)$ early in the process produces a higher increase of $r_{ix}$ compared to adding the link later. We also show in Lemma 2 below that $r_{ix}$ is not decreasing after adding $(u, x)$, which is intuitively clear. We now conclude from (7) that $\frac{\pi_x}{z_{xx}}$ is a submodular and nondecreasing function since $\frac{\pi_x}{z_{xx}}$ is a sum of submodular and nondecreasing terms.

When we greedily maximize a nonnegative nondecreasing submodular function we will always obtain a solution within a fraction $1 - \frac{1}{e}$ from the optimal according to [7] by Nemhauser *et al.* We now have that:

$$\frac{\pi_x^G}{z_{xx}^G} \geq \frac{\pi_x^o}{z_{xx}^o}(1 - \frac{1}{e}) \ .$$

Finally, we use the fact that $z_{xx}^G$ and $z_{xx}^o$ are numbers between 1 and $\frac{1}{1-\alpha^2}$.
□

For $\alpha = 0.85$ this gives an upper bound of $\frac{\pi_x^o}{\pi_x^G}$ of approximately 5.7 *It must be stressed that this upper bound is considerably smaller if $z_{xx}$ is close to the optimal value prior to the modification – if $z_{xx}$ cannot be improved then the upper bound is $\frac{e}{e-1} = 1.58$.* It *may* be the case that we obtain a bigger value of $\pi_x$ by greedily maximizing $\pi_x$ instead of $\frac{\pi_x}{z_{xx}}$, but $\pi_x$ (the PageRank of the target node throughout the Link Building process) is *not* a submodular function of $X$ so we cannot use the approach above to analyze this situation. To see that $\pi_x$ is not submodular we just have to observe that adding a backlink from a sink node creating a short cycle late in the process will produce a higher increase in $\pi_x$ compared to adding the link early in the process.

**Proof of Submodularity and Monotonicity of $r_{ix}$** Let $f_i(X)$ denote the value of $r_{ix}$ in $G(V, E \cup (X \times \{x\}))$ – the graph obtained after adding links from all nodes in $X$ to $x$.

**Lemma 1.** *$f_i$ is submodular for every $i \in V$.*

*Proof.* Let $f_i^r(X)$ denote the probability of reaching $x$ from $i$ without zapping, in $r$ steps or less, in $G(V, E \cup (X \times \{x\}))$. We will show by induction in $r$ that $f_i^r$ is submodular. We will show the following for arbitrary $A \subset B$ and $y \notin B$:

$$f_i^r(B \cup \{y\}) - f_i^r(B) \leq f_i^r(A \cup \{y\}) - f_i^r(A) \ . \tag{8}$$

We start with the induction basis $r = 1$. It is not hard to show that the two sides of (8) are equal for $r = 1$. For the induction step; if you want to reach $x$ in $r + 1$ steps or less you have to follow one of the links to your neighbors and reach $x$ in $r$ steps or less from the neighbor:

$$f_i^{r+1}(X) = \frac{\alpha}{outdeg(i)} \sum_{j: i \to j} f_j^r(X) \ . \tag{9}$$

where $j : i \to j$ denotes the nodes that $i$ links to – this set includes $x$ if $i \in X$. The outdegree of $i$ is also dependent on $X$. If $i$ is a sink in $G(V, E \cup (X \times \{x\}))$ then we can use (9) with $outdeg(i) = n$ and $j : i \to j = V$ – as explained in

Sect. 2, the sinks can be thought of as linking to all nodes in the graph. Please also note that $f_x^r(X) = 1$.

We will now show that the following holds for every $i \in V$ assuming that (8) holds for every $i \in V$:

$$f_i^{r+1}(B \cup \{y\}) - f_i^{r+1}(B) \leq f_i^{r+1}(A \cup \{y\}) - f_i^{r+1}(A) \ . \tag{10}$$

1. $i \in A$: The set $j : i \to j$ and $outdeg(i)$ are the same for all four terms in (10). We use (9) and the induction hypothesis to see that (10) holds.
2. $i \in B \setminus A$ :
   (a) $i$ is a sink in $G(V, E)$: The left hand side of (10) is 0 while the right hand side is positive or 0 according to Lemma 2 below.
   (b) $i$ is not a sink in $G(V, E)$: In this case $j : i \to j$ includes $x$ on the left hand side of (10) but not on the right hand side – the only difference between the two sets – and $outdeg(i)$ is one bigger on the left hand side. We now use (9), the induction hypothesis and $\forall X : f_x^r(X) = 1$.
3. $i = y$: We rearrange (10) such that the two terms including $y$ are the only terms on the left hand side. We now use the same approach as for the case $i \in B \setminus A$.
4. $i \in V \setminus (B \cup \{y\})$: As the case $i \in A$.

Finally, we use $\lim_{r \to \infty} f_i^r(X) = f_i(X)$ to prove that (8) holds for $f_i$. □

**Lemma 2.** $f_i$ *is nondecreasing for every* $i \in V$.

*Proof.* We shall prove the following by induction in $r$ for $y \notin B$:

$$f_i^r(B \cup \{y\}) \geq f_i^r(B) \ . \tag{11}$$

We start with the induction basis $r = 1$.

1. $i = y$: The left hand side is $\frac{\alpha}{outdeg(y)}$ where $outdeg(y)$ is the new outdegree of $y$ and the right hand side is at most $\frac{\alpha}{n}$ (if $y$ is a sink in $G(V, E)$).
2. $i \neq y$: The two sides are the same.

For the induction step; assume that (11) holds for $r$ and all $i \in V$. We will show that the following holds:

$$f_i^{r+1}(B \cup \{y\}) \geq f_i^{r+1}(B) \ . \tag{12}$$

1. $i = y$:
   (a) $i$ is a sink in $G(V, E)$: The left hand side of (12) is $\alpha$ and the right hand side is smaller than $\alpha$.
   (b) $i$ is not a sink in $G(V, E)$: We use (9) in (12) and obtain simple averages on both sides with bigger numbers on the left hand side due to the induction hypothesis.
2. $i \neq y$: Again we can obtain averages where the numbers are bigger on the left hand side due to the induction hypothesis.

Again we use $\lim_{r \to \infty} f_i^r(X) = f_i(X)$ to conclude that (11) holds for $f_i$. □

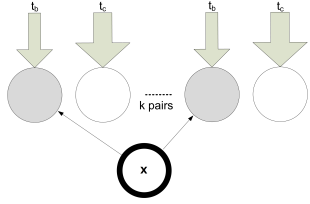

**Fig. 4.** Sink versus sink for the r-Greedy. Not shown is a large isolated clique of size $t_i$.

### 3.3 Lower Bound for r-Greedy

Let
$$r_x = 1 + \sum_{i \neq x} r_{ix} = \pi_x / z_{xx} \ .$$

be the *reachability* of node $x$. The r-Greedy algorithm above improves the reachability of the target node by the maximal amount at each step. However, it misses opportunities to improve $\pi_x$ by increasing the $z_{xx}$ value (short loops), and so it is quite easy to show that there is an infinite graph family where the approximation ratio is
$$\frac{\textit{max possible } z_{xx}}{\textit{min possible } z_{xx}} = \frac{1/\left(1-\alpha^2\right)}{1} = \frac{1}{1-\alpha^2}$$
which is equal to about 3.6 for $\alpha = 0.85$. In order to force a greater approximation ratio, we would have to consider graph families that use the independent set aspect of link building, as discussed in [8].

**Theorem 3.** *Consider the Link Building problem with target node $x$. Let $G = (V, E)$ be some directed graph. Let $\pi_x^o$ denote the highest possible PageRank that the target node can achieve after adding $k$ links, and $\pi_x^o$ denote the PageRank after adding the links returned by the r-Greedy algorithm above. Then for any $\epsilon > 0$ there exist infinitely many different graphs $G$ where*

$$\frac{\pi_x^o}{\pi_x^r} > \frac{1}{1-\alpha^2} - \epsilon \ . \tag{13}$$

*Proof.* Consider the input given by Fig. 4 for some parameter $t_b$ and $t_c = t_b + 1$. At the beginning, $r_x \approx 1$ since it is unreachable, except by zapping, from any other node in the graph (the large isolated clique ensures that this zapping traffic is insignificant). Throughout the link building process, adding a link from a shaded node will add $\alpha + t_b \, \alpha^2$ to $r_x$, while adding a link from a light node will add $\alpha + (t_b + 1)\, \alpha^2$ (minus some insignificant zapping traffic that we lose due to these nodes no longer being sinks). Hence, r-Greedy will always choose to add links from the light nodes and ignore the possibility of creating short loops.

Now we analyse the effect of the links on $\pi_x$ as we have in Sect. 3.1. First we consider the graph before any additional links are added. Again due to symmetry the $k$ shaded nodes will have the PageRank $\pi_b$, and the light nodes $\pi_c$. All tail nodes have no incoming links and, due to symmetry, will have the same PageRank denoted by $\pi_t$. The PageRank of the target node is $\pi_x$ and the PageRank of each node in the isolated clique is $\pi_i$.

Again we compute the initial PageRanks for this kind of symmetric graph by writing a linear system of equations based on the identity $\pi^T = \pi^T Q$. The total number of nodes here is $n = k\,(t_c + t_b + 2) + t_i + 1$.

$$\pi_t = \frac{1-\alpha}{n} + \frac{k\,\alpha\,(\pi_b + \pi_c)}{n}$$
$$\pi_x = \pi_t$$
$$\pi_b = \pi_t + \alpha\left(\frac{\pi_x}{k} + t_b \pi_t\right)$$
$$\pi_c = \pi_t + \alpha\, t_c \pi_t$$
$$\pi_i = \pi_t + \alpha \pi_i = \frac{\pi_t}{1-\alpha}\;.$$

The greedy will choose $k$ links from the light nodes as proven above (if $t_c > t_b$) giving:

$$\pi_t^r = \frac{1-\alpha}{n} + \frac{k\,\alpha\,\pi_b^r}{n}$$
$$\pi_x^r = \pi_t^r + k\,\alpha\,\pi_c^r$$
$$\pi_b^r = \pi_t^r + \alpha\left(\frac{\pi_x^r}{k} + t_b \pi_t^r\right)$$
$$\pi_c^r = \pi_t^r + \alpha\, t_c \pi_t^r$$
$$\pi_i^r = \frac{\pi_t^r}{1-\alpha}\;.$$

The optimal solution is to choose $k$ links from the shaded nodes with a resulting PageRank vector described by the following system:

$$\pi_t^o = \frac{1-\alpha}{n} + \frac{k\,\alpha\,\pi_c^o}{n}$$
$$\pi_x^o = \pi_t^o + k\,\alpha\,\pi_b^o$$
$$\pi_b^o = \pi_t^o + \alpha\left(\frac{\pi_x^o}{k} + t_b \pi_t^o\right)$$
$$\pi_c^o = \pi_t^o + \alpha\, t_c \pi_t^o$$
$$\pi_i^o = \frac{\pi_t^o}{1-\alpha}\;.$$

We now parameterize our variables simply thus:

$$t_b = c$$
$$t_c = c+1$$
$$t_i = c^2\;.$$

So that a single variable determines the size of the graph. We solve these systems and calculate the approximation ratio of the naive algorithm, giving an expression with numerator

$$-((a^2ck + ak + 1)((1 - a^4)(c+1)k + (1 - a^2)ck +$$
$$(-a^3 - a + 2)k + c^2 - a^2 + 1))$$

and denominator

$$((-a^3 - a^2 + a + 1)(c+1)k + (a+1)ck + (-a^2 + a + 2)k +$$
$$(a+1)c^2 + a + 1)((a^3 - a^2)(c+1)k + (a^2 - a)k + a - 1)$$

Letting the size of the graph go to infinity, we get the following limit. We note that this limit is approached from the left.

$$lim_{c \to \infty} \frac{\pi_x^o}{\pi_x^r} = \frac{1}{1 - \alpha^2} \quad .$$

And the theorem statement follows from this.

□

## 4 Discussion and Open Problems

We have presented a constant-factor approximation polynomial time algorithm for Link Building. We also presented a lower bound for the approximation ratio achieved by a perhaps more intuitive and simpler greedy algorithm. The problem of developing a polynomial time approximation scheme (PTAS) for Link Building remains open.

## References

1. Konstantin Avrachenkov and Nelly Litvak. The Effect of New Links on Google Pagerank. *Stochastic Models*, 22(2):319–331, July 2006.
2. Monica Bianchini, Marco Gori, and Franco Scarselli. Inside pagerank. *ACM Transactions on Internet Technology*, 5(1):92–128, February 2005.
3. Sergey Brin and Lawrence Page. The Anatomy of a Large-Scale Hypertextual Web Search Engine. *Computer networks and ISDN systems*, 30(1-7):107–117, 1998.
4. C Dekerchove, L Ninove, and P Vandooren. Maximizing PageRank via outlinks. *Linear Algebra and its Applications*, 429(5-6):1254–1276, 2008.
5. Rand Fishkin. Search Engine Ranking Factors, 2011.
6. Amy N A.N. Langville and Carl D C.D. Meyer. Deeper inside pagerank. *Internet Mathematics*, 1(3):335–380, 2004.
7. G.L. L. Nemhauser, L.A. A. Wolsey, and M.L. L. Fisher. An analysis of approximations for maximizing submodular set functions. *Mathematical Programming*, 14(1):265–294, December 1978.
8. Martin Olsen. Maximizing PageRank with New Backlinks. In *Algorithms and complexity (CIAC)*, pages 37–48, 2010.
9. Martin Olsen, Anastasios Viglas, and Ilia Zvedeniouk. A Constant-Factor Approximation Algorithm for the link building Problem. In *Combinatorial Optimization and Application*, volume 65, pages 87–96. Springer Berlin Heidelberg, August 2010.